\documentclass[journal=jacsat,manuscript=article,layout=twocolumn]{achemso}
\usepackage{xcolor}
\usepackage{lipsum} % Required to insert dummy text
\usepackage[version=4]{mhchem}
\usepackage{siunitx}
\DeclareSIUnit\Molar{M}
\usepackage[italic]{mathastext}
\usepackage{mhchem}
\usepackage{multirow}
\usepackage{multicol}
\usepackage{xcolor}

\usepackage{orcidlink}

\usepackage[colorinlistoftodos]{todonotes}
\usepackage{enumitem} % better handling
\usepackage{pifont} % some unicode-eqsue characters

\usepackage{newpxtext}

\makeatletter
\newcommand{\ie}{\emph{i.e.}\@ifnextchar.{\!\@gobble}{}}
\newcommand{\eg}{\emph{e.g.}\@ifnextchar.{\!\@gobble}{}}
\usepackage{xr}
\makeatletter
\newcommand*{\addFileDependency}[1]{% argument=file name and extension
  \typeout{(#1)}
  \@addtofilelist{#1}
  \IfFileExists{#1}{}{\typeout{No file #1.}}
}
\makeatother
\newcommand*{\myexternaldocument}[1]{%
    \externaldocument{#1}%
    \addFileDependency{#1.tex}%
    \addFileDependency{#1.aux}%
}
\myexternaldocument{supplementary}

\usepackage{hyperref}
\hypersetup{colorlinks=true,linkcolor = blue,citecolor=blue}

\author{Tarun Maity}
\author{Yogendra Kumar}
\affiliation[IISc]
{Center for Condensed Matter Theory, Department of Physics, Indian Institute of Science, Bangalore 560012, India}
\author{Ashish Kumar Singha Deb}
\author{Sk Musharaf Ali}
\affiliation[BARC]
{Chemical Engineering Division, Bhabha Atomic Research Centre, Mumbai, India}
\author{Prabal K Maiti}
\email{maiti@iisc.ac.in}
\affiliation[IISc]
{Center for Condensed Matter Theory, Department of Physics, Indian Institute of Science, Bangalore 560012, India}
\title[An \textsf{achemso} demo]
  {Enhanced and Efficient Extraction of Uranyl Ions from Aqueous Waste through Graphene/CNT-PAMAM Nanocomposites }

\abbreviations{IR,NMR,UV}
\keywords{American Chemical Society, \LaTeX}

\begin{document}
%%%%%%%%%%%%%%%%%%%%%%%%%%%%%%%%%%%%%%%%%%%%%%%%%%%%%%%%%%%%%%%%%%%%%
\maketitle

\twocolumn[
\begin{@twocolumnfalse}
\begin{abstract}
The increasing threat of uranium contamination to environmental and human health due to its radiotoxicity demands the development of novel and efficient adsorbents for remediation.
In this study, we investigated the potential of poly(amidoamine) (PAMAM) dendrimers of generations 1 to 4 (G1 - G4)  functionalized with graphene and carbon nanotubes (CNTs) as adsorbents for uranyl ion removal from aqueous solutions. 
By combining atomistic molecular dynamics (MD) simulations with experimental validation, we examined the influence of pH, uranyl ion concentration, and dendrimer generation on adsorption behavior.
Our study revealed that uranyl ion adsorption is greater when PAMAM is grafted onto graphene/CNT than pristine PAMAM.
However, PAMAM-grafted CNTs exhibit superior adsorption capacity at specific uranyl concentrations due to their curvature and abundant accessible binding sites. Higher-generation PAMAM dendrimers grafted onto graphene/CNTs exhibit greater adsorption capacity due to the increased availability of binding sites, which is consistent with experimental observations. The adsorption capability of uranyl ions in all four generations of the PAMAM dendrimer increased as the concentration of uranyl ions increased.
Adsorption capacity increases with increasing uranyl ion concentration, and adsorption occurs on both PAMAM and graphene/CNT surfaces, with saturation observed at higher concentrations. 
This study provides insights into the adsorption mechanisms and highlights the potential of PAMAM-based nanocomposites for efficient uranyl ion extraction and environmental remediation.
\end{abstract}
%\section{Keywords}
%PAMAM dendrimers $;$ graphene/CNT-PAMAM nanocomposites  $;$ Uranyl ions $;$ Adsorption $;$ waste water $;$ MD simulation $;$ ultrafiltration.
% \section{Synopsis}
% This study demonstrates that graphene/CNT-PAMAM nanocomposites as efficient, sustainable adsorbents for uranyl ion remediation, offering superior performance under acidic conditions.\\
 \end{@twocolumnfalse}
]

%%%%%%%%%%%%%%%%%%%%%%%%%%%%%%%%%%%%%%%%%%%%%%%%%%%%%%%%%%%%%%%%%%%%%
%% Start the central part of the manuscript here.
%%%%%%%%%%%%%%%%%%%%%%%%%%%%%%%%%%%%%%%%%%%%%%%%%%%%%%%%%%%%%%%%%%%%%
\newpage 
\section{Introduction} 
{\label{sec:level1}}
In this era of rapidly increasing energy demand and depreciating fossil fuel-based energy sources, the growth of nuclear power has become a pivotal focus of the global energy landscape\cite{international2019nuclear}. Uranium (\ce{U (VI)}), the essential fuel for nuclear reactors, is primarily sourced from land-based ores \cite{xie2023uranium}. However, due to the limited availability of these traditional sources, there is growing interest in extracting uranium from unconventional sources, such as waste coal ash and seawater \cite{xie2023uranium}. 
Although this diversification of uranium sources holds 
promise for sustaining nuclear energy, it also raises significant environmental and health concerns due to radioactive, carcinogenic, and toxic properties of uranium\cite{manaka2008uranium,arfsten2001review}.
The toxicity of uranium and its compounds can result in progressive or irreversible renal injury and in acute cases may lead to kidney failure and even death \cite{liu2012high}.It has tendency to accumulate bone and causes damages\cite{semiao2010impact}.
 
Uranium concentrations in shallow groundwater across the world vary widely, from \num{0.0} to \SI{0.133}{\milli\gram}, often exceeding the World Health Organization's (WHO) recommended limit of \SI{0.0030}{\milli\gram\per\liter} for drinking water\cite{world2008guidelines}.
The radioactive liquid waste generated during front end and back end processes of nuclear fuel cycle contain significant concentration of various radionuclides including uranium\cite{van2004separation, raj2006radioactive}.  The permissible discharge level of uranium for nuclear industries effluent ranges from 0.05 to \SI{0.5}{\milli\gram\per\liter}\cite{anirudhan2009improved, hu2019study}.
Therefore, extraction of uranium \ce{U(VI)} from aqueous solutions is crucial for environmental preservation and the sustainable development of nuclear energy.

In the aqueous solution, uranium  (\ce{U(VI)}) is primarily found in highly soluble and mobile form, uranyl cation (\ce{UO2^2+}). \cite{alessi2012quantitative,wang2013mobile}
 Therefore, various methods within the nuclear industry, such as chemical precipitation \cite{kim2009precipitation,foster2020uranyl,tabushi1979extraction}, solvent extraction \cite{preston1995solvent,kumar2011brief,kim2013recovery,georgiou2023uranium}, reverse osmosis \cite{kucera2023reverse,hsiue1989treatment,sorg1990removal,lin1987treatment}, ion exchange \cite{nachod2012ion,rodrigues2012ion,arden2012water}, and flotation \cite{matis1994flotation}, are employed to treat aqueous nuclear waste effectively. However, these conventional methods have significant drawbacks in producing secondary pollutants, as well as high costs and energy requirements. Physical adsorption \cite{ruthven1984principles} involving only relatively weak intermolecular forces than chemisorption is one of the most attractive and efficient way to remove uranium from aqueous solutions since it is inexpensive, easy to use, and very successful \cite{zhang2015adsorption,lei2022efficient,hua2012heavy,tripathi2013uranium,simsek2022high}. Several traditional adsorbents such as polymer fibers \cite{kumar2023efficacy,ismail2020hydrophilic,gleissner2022surface}, silica gels \cite{koner2010cationic,katsumata2003removal}, and carbon-based materials \cite{duan2020removal,qasem2021removal} \textit{e.g.} carbon nanotube (CNT)\cite{xu2008removal,yu2016removal,druchok2017carbon,wang2020efficient}, graphene (GRA) \cite{ahmad2020adsorptive}, graphene-oxide (GO) \cite{ai2018effect,lan2023rigidity,liu2019graphene,peng2017review,wang2013adsorption,sherlala2019adsorption}, and porous activated carbon \cite{ao2018microwave,wang2020cr} has been used for the recovery of uranium ions and heavy metals from the aqueous solution. 
 
 Dendrimer, a class of hyperbranched polymers with large pore size distributions, globular shapes, and tunable surface chemistries, have found extensive use in nanotechnology and biomedicine \cite{pramanik2021dendrimers, svenson2012dendrimers,deirram2019ph,nandy2011dna, maity2022molecular, sajid2018removal,viltres2021polyamidoamine}. Recently, their potential for chelating heavy metal ions from nuclear waste and seawater has gained significant attention due to their high capacity, selectivity, and recyclability \cite{diallo2008dendritic, sajid2018removal,viltres2021polyamidoamine,maity2023efficient,maiti2004structure,maiti2017pamam,sengupta2017amidoamine,kumar2017poly,kumar2016sorption,kumar2017understanding,ilaiyaraja2013removal,ilaiyaraja2017surface}. Diallo et al. reported results of an experimental investigation of the binding of Cu(II) ions to poly(amidoamine) (PAMAM) dendrimers in aqueous solutions \cite{diallo1999poly}. They further reported the feasibility of using dendrimer enhanced ultrafiltration (DEUF) to recover Cu(II) from aqueous solutions \cite{diallo2005dendrimer}. 
 Later, Xu et al. studied the PAMAM dendrimers of various generations and terminal functional groups for removal of copper(II) in a sandy soil \cite{xu2005removal}. 
 %Wang et al. reported the results of batch experiments of U(VI) adsorption to synthetic and biogenic MnO$_2$ \cite{wang2013mobile}. 
 Defever et al. reported the results from experiments and atomistic molecular dynamics simulations on the remediation of naphthalene by polyamidoamine (PAMAM) dendrimers and graphene oxide (GrO)\cite{defever2015pamam}.
Sun et al. reported the adsorption and desorption of U(VI) on graphene oxides (GOs), carboxylated GOs (HOOC-GOs), and reduced GOs (rGOs) by batch experiments, EXAFS technique, and computational theoretical calculations \cite{sun2015adsorption}.
Lan et al. reported the adsorption of uranyl on graphene oxide (GO)  in collaboration with humic acid (HA) using molecular dynamics simulations\cite{lan2019competition}.
%Kotte et al. studied the Cu(II) recovery from aqueous solutions using mixed matrix polyvinylidene ﬂuoride (PVDF) membranes with in situ synthesized PAMAM particles \cite{kotte2015mixed}.
Li et al. reported the adsorption capacity of uranyl ions using amidoxime-functionalized carboxymethyl $\beta$-cyclodextrin/graphene aerogel (GDC) \cite{li2021high}.
%Previous experimental\cite{hayati2016synthesis,hayati2017super,hayati2018heavy,guo2022adsorption} and simulation\cite{kommu2017removal,maity2023efficient} studies have utilized PAMAM grafted with carbonaceous material to remove heavy metals from wastewater in an efficient manner \cite{viltres2021polyamidoamine,guo2022adsorption}. 
%The articles reported either have shown the importance of dendrimers or GO or functionalized GO for removal of Copper ions or uranyl ions. There was no study on PAMAM dendrimer functionalized GO for uranyl ion adsorption either by experiments or MD simulations. 
%These findings imply that the PAMAM dendrimer grafted with carbonaceous material may enhance the effective extraction of heavy metal ions from sea/wastewater. 
%However, there is a lack of thorough understanding regarding the binding behaviour of heavy metal ions to such nanocomposites. 
%To the best of our knowledge this is the first study by combining MD simulations and experiments to develop and highly efficient adsorbents for uranyl ion by coupling PAMAM dendrimers with GO. 
Despite these advancements, previous studies have predominantly focused on either dendrimers or graphene-based materials for removing heavy metal ions, with limited exploration of nanocomposites that integrate both materials. Recent experimental and simulation studies have shown that PAMAM dendrimers grafted onto carbonaceous materials can enhance the removal efficiency of heavy metals from wastewater \cite{hayati2016synthesis, hayati2017super, hayati2018heavy, guo2022adsorption, kommu2017removal, maity2023efficient}. However, no studies have investigated PAMAM dendrimer-functionalized graphene oxide (GO)/CNT for uranyl ion adsorption through combined experimental and MD simulation approaches.
Also, there is a lack of thorough understanding regarding the binding behaviour of heavy metal ions to such nanocomposites.

 This study focuses on developing a novel nanocomposite adsorbent for extracting uranyl ions from aqueous solutions using graphene and CNTs nanocomposites functionalized with PAMAM dendrimers. Through atomistic molecular dynamics (MD) simulations and and experimental studies, we systematically evaluate the influence of pH and uranyl ion concentration on the adsorption process. Experimental evaluation of the mechanism of interaction between the complex structure of higher generation PAMAM dendrimer grafted CNT and GO with uranyl metal ions is very difficult owing to the presence of multiple number of binding sites in three dimensional network.
Our findings shed light on the underlying adsorption mechanisms and demonstrate the efficacy of PAMAM-based nanocomposites for selective and efficient uranium extraction.

The rest of the paper is organized as follows: Section II details the model-building and simulation methodologies employed. Section III presents the results of molecular dynamics (MD) simulations and experimental studies. Finally, Section IV summarizes the key findings and discusses their implications for future research directions.

\section{Modelling and Simulation Details:} \label{sec:level2}
\subsection{Model Building:}
\begin{figure}[!h]
     \centering
     \includegraphics[width=\linewidth]{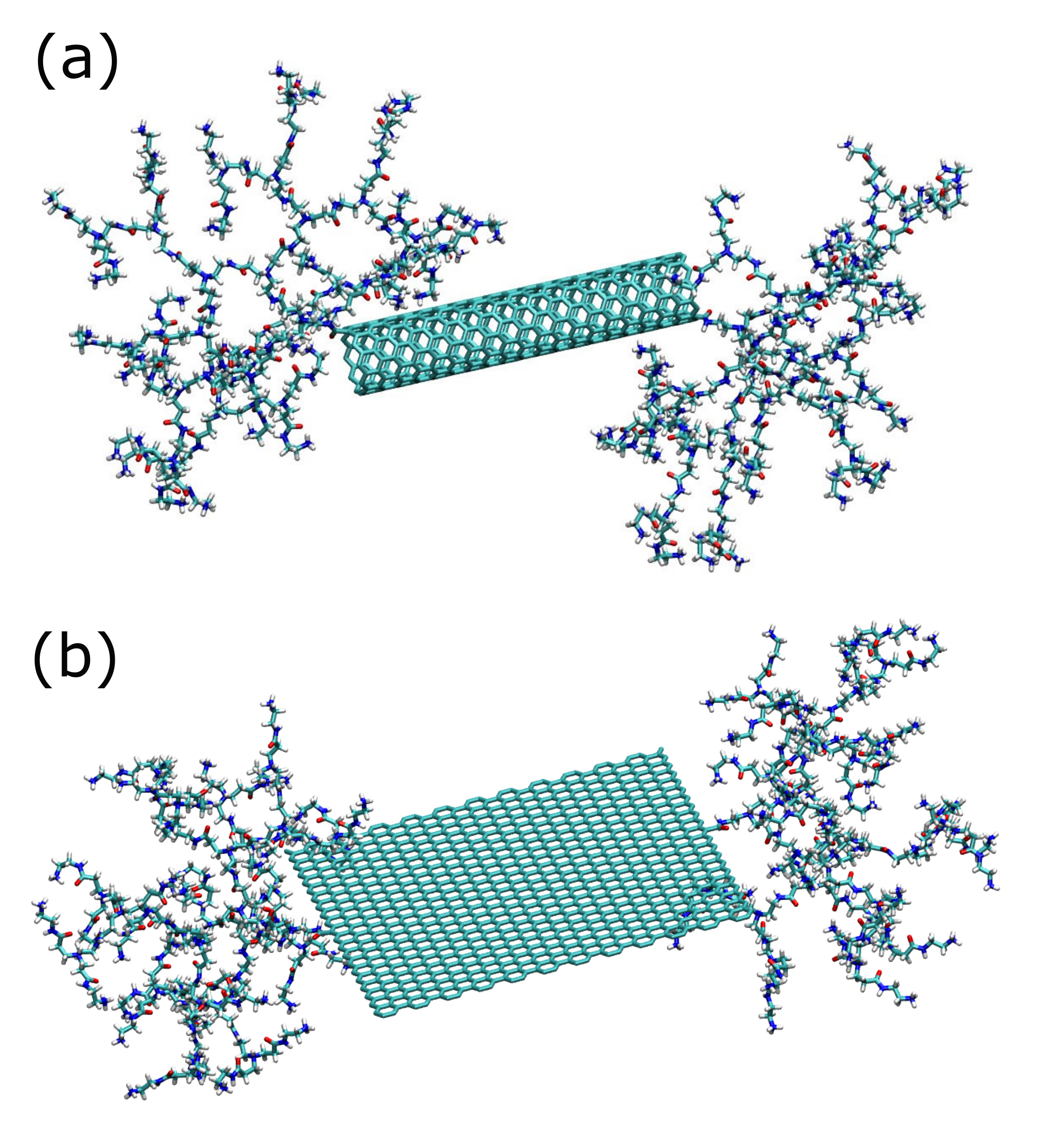}
     \caption[Initial snapshots of the graphene-PAMAM and CNT-PAMAM nanocomposite.]{Initial snapshots of the (a) graphene-PAMAM and (b) CNT-PAMAM nanocomposite. The grafting-from approach has been used to make a covalent bond between the graphene/CNT and dendrimer.}
     \label{fig:graft}
 \end{figure}

To prepare the graphene-PAMAM and CNT-PAMAM nanocomposites, we first built graphene sheets of dimension ($\SI{5}{\nano\meter} \times \SI{5}{\nano\meter}$ ) and single-walled carbon nanotube (SWCNT) of chirality ($10,0$), using visual molecular dynamics (VMD) software\cite{humphrey1996vmd}. Polyamidoamine (PAMAM) dendrimers of generations 1 to 4 (G1-G4) at various protonation levels (corresponding to different pH conditions) were modeled using the dendrimer builder toolkit (DBT)\cite{maingi2012dendrimer}. 
Two protonation states were employed to represent the two distinct pH conditions observed in acid-base titration  experiments\cite{van1998acid}. At neutral pH (pH $\approx 7$ ), the primary amines are protonated; at low pH (pH $<$ 3), both primary and tertiary amines get protonated \cite{maiti2004structure, van1998acid, maiti2005effect}.
The \textit{xLEaP} module of AMBER\cite{case2021amber} was used to make covalent bonds between the edge of graphene/CNT and the dendrimer's core through the grafting-from approach via a carboxylic group (\ce{-COOH})\cite{gosika2020covalent}  as shown in Figure \ref{fig:graft}.
This covalent bond alters the chemical environment for the core atoms, including the ethylenediamine (EDA) core and the grafting sites of the graphene/CNT. To account for these changes, the atomic charges of the core atoms and the attached hexagonal ring of graphene/CNT were recalculated.
This calculation employed the electrostatic potential (ESP) method implemented in Gaussian 09 software \cite{gaussian0920091} by using the Hartree-Fock (HF) theory and the 6-31G(d) basis set. Subsequently, the partial charge of each atom was calculated through the restrained electrostatic potential (RESP) using the antechamber module of AMBER20\cite{case2021amber}. Partial atomistic charges of the monomer units and the terminal groups were adopted from our previous work \cite{maingi2012dendrimer, gosika2020covalent}.

Next, graphene/CNT-PAMAM dendrimer nanocomposite was solvated using a box of TIP3P  ~\cite{jorgensen1983comparison, price2004modified} water.
We ensured \SI{15}{\angstrom} water layer in all three directions from the surface of the nanocomposite.
%We used different protonation states of the dendrimer corresponding to neutral and low pH.
We added Uranyl (\ce{UO_2^{2+}}) ions to the system to maintain the molarity at 0.05, 0.1, 0.2, 0.3, 0.4, 0.5, 0.6, 0.7, 0.8, 0.9, and 1.0 M for low pH and neutral pH solutions. Table \ref{tab:systemDetails} gives the simulated systems' details. 
To maintain charge neutrality of the simulation box at both pH conditions, nitrate (\ce{NO3^{-}}) counterions were added. The details of the total number of uranyl and nitrate ions at different pH and uranyl ion concentrations are shown in Table \ref{tab:systemDetails}.

\subsection{Simulation Details:}
The General Amber Force Field (GAFF) \cite{wang2004development} was employed to model intra- and intermolecular interactions involving dendrimers and graphene/CNTs since it has been validated in the previous work from our research group\cite{gosika2020covalent}.
The bonded and nonbonded parameters for uranyl and nitrate ions were adopted from Guilbaud and co-workers \cite{guilbaud1993hydration}.

\begin{figure*}[h!]
    \centering
    \includegraphics[width=1\linewidth]{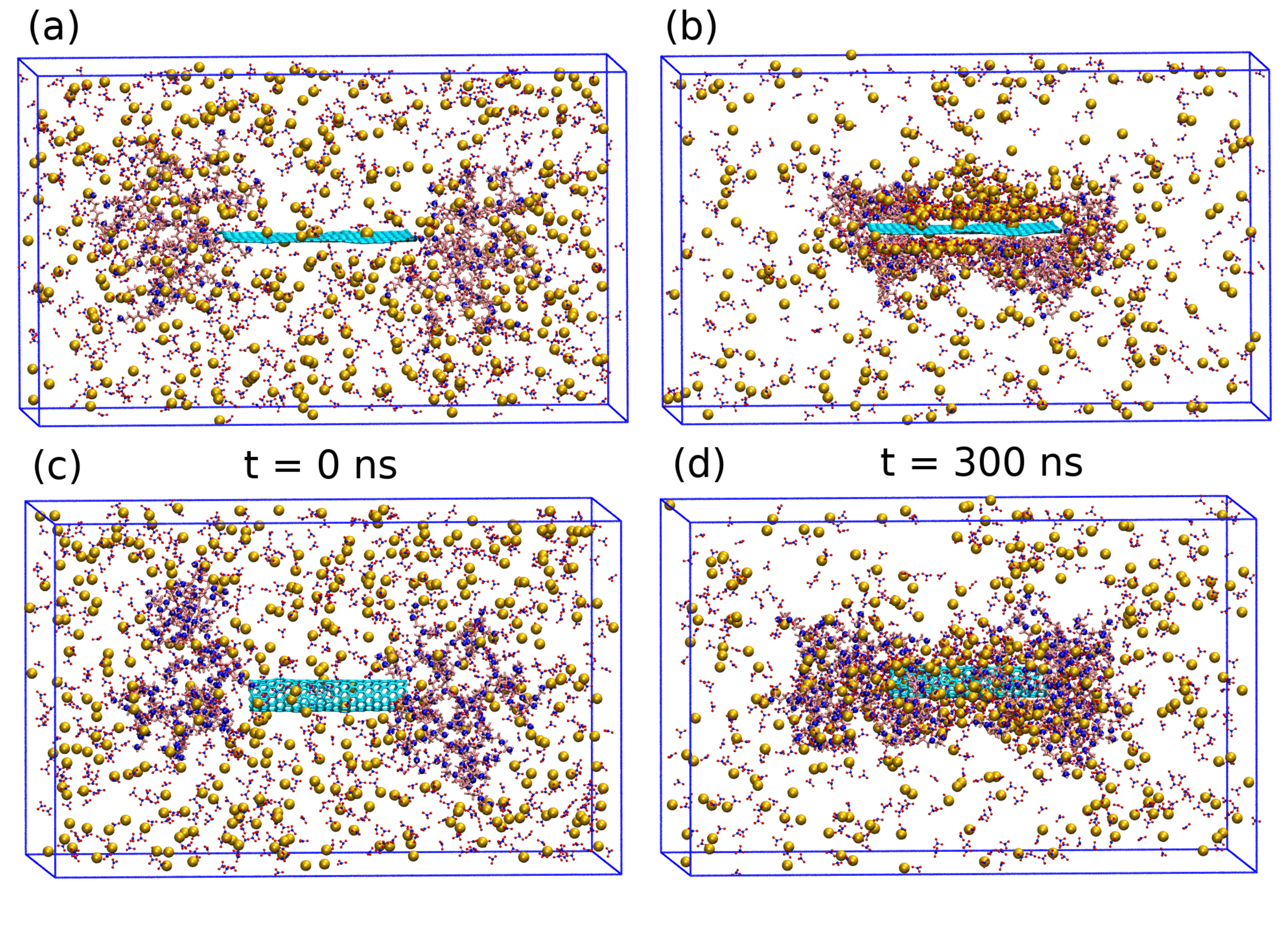}
    \caption[Instantaneous snapshots of uranyl ion adsorption on PAMAM-functionalized graphene and carbon nanotube (CNT) nanocomposites ]{Instantaneous snapshots of uranyl ion adsorption on PAMAM-functionalized (a,b) graphene and (c,d) carbon nanotube (CNT) nanocomposites at low pH and $0.5$ M uranyl ion concentration. The snapshots are taken at the simulation's beginning (t = 0 ns) and end (t = 300 ns). Color code: graphene and CNT are in cyan, PAMAM dendrimer in pink, uranyl ions in orange (bead form), and nitrate ions in CPK representation (nitrogen in blue, oxygen in red). Water molecules are omitted for clarity.}
    \label{fig:snapshot}
\end{figure*}

All the simulations were performed using the PMEMD and PMEMD CUDA modules of AMBER21 \cite{case2021amber}. To remove initial bad contact, each solvated graphene/CNT-dendrimer nanocomposite is subjected to \num{1000} steps of steepest descent followed by \num{1000} steps of conjugate gradient energy minimization.
Subsequently, the systems were gradually heated in the NVT ensemble to relax to the temperature of interest, \SI{300}{\kelvin} using a Langevin thermostat \cite{schneider1978molecular} with a friction collision frequency of \SI{1}{\per\pico\second}. 
This heating procedure consists of four successive stages: 
%initially, through an MD simulation, 
the temperature was increased from \SI{10}{\kelvin} to \SI{50}{\kelvin} over \SI{6}{\pico\second}; then, from \SI{50}{\kelvin} to \SI{100}{\kelvin} over \SI{12}{\pico\second}; followed by a third stage from \SI{100}{\kelvin} to \SI{200}{\kelvin} over \SI{10}{\pico\second}
; and finally, from \SI{200}{\kelvin} to \SI{300}{\kelvin} over an additional \SI{12}{\pico\second}. 
During heating, solute \emph{i.e.} graphene/CNT-PAMAM nanocomposite was restrained using a harmonic spring with a force constant of 20 kcal mol\textsuperscript{-1} \AA\,\textsuperscript{-2}. 
Subsequently, MD simulations were performed at an NPT ensemble for \SI{5}{\nano\second} to achieve the equilibrated density. The Berendsen barostat and Langevin thermostat were used to control the pressure at 1 bar and temperature at \SI{300}{\kelvin} with pressure coupling constant of \SI{0.5}{\per\pico\second} and temperature coupling constants of \SI{1}{\per\pico\second}, respectively. 
Finally, we performed \SI{300}{\nano\second} long NVT production run at \SI{300}{\kelvin}, using a Langevin thermostat. 
The long-range electrostatic interactions were calculated using the particle mesh Ewald (PME) method ~\cite{darden1993particle} with a real space cut off of \SI{9}{\angstrom} and a reciprocal space convergence tolerance of $5 \times 10^{-4}$ \SI{}{\angstrom}.
To facilitate larger timesteps (\SI{2}{\femto\second}), we constrained all bonds involving hydrogen atoms and the H-O-H angle in water to their equilibrium positions using the SHAKE algorithm with a \SI{0.005}{\angstrom} tolerance.
Finally, all the system’s properties were obtained by calculating the averages over the last \SI{20}{\nano\second} of the simulation trajectory 

\section{Experimental Details:}
\subsection{Synthesis of CNT-PAMAM and GO-PAMAM up to generation 4}
PAMAM dendrons starting from generation zero (G0) to generation four (G4) were grown on the surface of oxidized multi-walled carbon nanotubes (MWCNTs) and graphene oxide (GO) through the `grafting from' method by a divergent synthesis route using a carboxylic acid group as a linker. 
% The details of the synthetic procedure for the preparation of CNT-PAMAM-G4 and GO-PAMAM-G4 from MWCNTs and GO, respectively, are given in the Supporting Information (SI).

\subsection{Batch adsorption studies of uranyl ions by synthesized CNT/GO-PAMAM}
The performances of uranyl ions (\ce{UO_2^{2+}}) adsorption by each full generation of CNT/GO-PAMAM up to generation 4 were evaluated in static batch mode. An aqueous solution containing uranyl ions adsorbate was equilibrated with the prepared adsorbents (full generations of CNT/GO-PAMAM from G1 to G4). The supernatant aliquot after the equilibration period was analyzed for its uranyl ion concentration, which was compared with the feed value, and the adsorption capacities of each generation of CNT/GO-PAMAM were calculated under different experimental conditions.
The details of the adsorption experiment and calculations are described in the SI.

%%%%%%%%%%%%%%%%%%%%%%%%%%%%%%%%%%%%%%%%%%%%%%%%%%%%%%%%%%%%%%%%%%%%%%%%%%%%%%%%%%%%%%%%%%%%%%%%%%%%%%%%%%%%%%%%
\section{Results and discussion}
%\subsection{Uranyl Ion Adsorption and Equilibrium}
We investigated the adsorption of uranyl ions onto graphene-PAMAM and CNT-PAMAM dendrimer nanocomposites across various generations of dendrimers, pH conditions, and initial uranyl ion concentrations.
A uranyl ion is considered to be adsorbed when located within the first solvation shell (\SI{8.5}{\angstrom} of PAMAM binding sites (amine, amide, and carbonyl groups) or the graphene/CNT surface \cite{maity2023efficient}.
Figure \ref{fig:snapshot} illustrates the adsorption of uranyl ions onto graphene/CNT-PAMAM nanocomposites at low pH and 0.5 M uranyl ion concentration. The snapshots are taken at the beginning ($t=0$ ns) and end ($t=300$ ns) of the simulation. Figure \ref{fig:snapshot} illustrates that initially, there is no adsorption, and as simulation time progresses, uranyl ions get adsorbed on the graphene/CNT-PAMAM nanocomposite. We computed the total number of uranyl ions adsorbed onto the nanocomposite to quantify the adsorption.

\begin{figure*}[!h]
    \centering
    \includegraphics[width=0.8\linewidth]{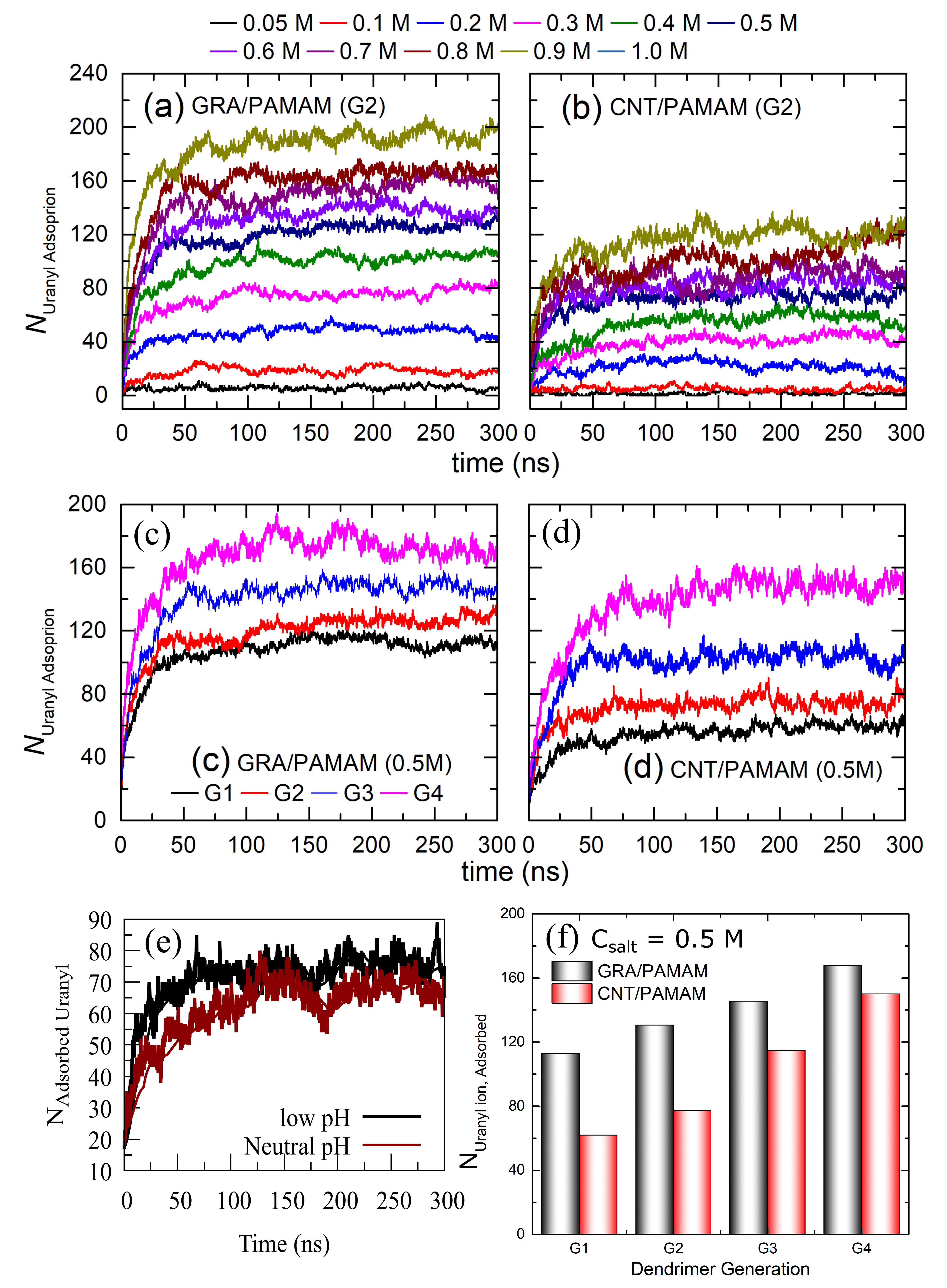}
    \caption[Temporal variation in the number of uranyl (\ce{UO2^2+}) ions]{Temporal variation in the number of uranyl (\ce{UO2^2+}) ions adsorbed by the (a-b) graphene/CNT-PAMAM (G2) nanocomposite as a function of uranyl ion concentration and (c-d) as a function of different dendrimer generation at 0.5 M concentration, (e) at low pH (black) and neutral pH (red) conditions, and the average of (c-d) is plotted in (f).}
    \label{fig:num_U_ads_combined}
\end{figure*}

Figure \ref{fig:num_U_ads_combined}  depicts that the number of adsorbed uranyl ions gradually increases over time and reaches equilibrium for all the uranyl ion concentrations, pH of the solution, and generation of the dendrimer. 
The number of adsorbed uranyl ions depends on dendrimer generation $(G1-G4)$, pH of the solution, and initial uranyl ion concentration ($C_0$). 
Figure \ref{fig:num_U_ads_combined}(a,b)) shows that increasing the initial concentration of uranyl ions leads to a higher number of ions adsorbed onto the nanocomposite, likely because more uranyl ions are available to interact with the adsorbent surface.
%Higher uranyl ion concentrations enhanced adsorption due to a higher probability of encountering the adsorbent surface (Figure \ref{fig:num_U_ads_combined}(a,b)).
 Figure \ref{fig:num_U_ads_combined}(c,d) shows the adsorption trends for different dendrimer generations at a fixed salt concentration of 0.5 M. The order of uranyl ion adsorption for both graphene-PAMAM and CNT-PAMAM nanocomposites follows G4 $>$ G3 $>$ G2 $>$ G1, reflecting an increase in binding sites with increasing dendrimer generation. 
Figure \ref{fig:num_U_ads_combined} (e) depicts more adsorbed uranyl ions in the CNT-PAMAM nanocomposite at low pH conditions compared to neutral pH. This pH dependency indicates that the protonation state of the PAMAM, modulated by the solution pH, influences the interaction of the nanocomposite with uranyl ions. 
To compare the number of adsorbed uranyl ions in nanocomposites, we calculated the average number of adsorbed uranyl ions ($< N_{Adsorbed Uranyl} >$) onto nanocomposites using the last \SI{20}{\nano\second} of the simulation trajectory.
Interestingly, Figure \ref{fig:num_U_ads_combined} (f) reveals higher average uranyl ion adsorption for graphene-PAMAM (G2) than for CNT-PAMAM (G2) at various uranyl ion concentrations. This observation is likely to be attributed to more carbon atoms in graphene (1008) than in CNT (800), providing more potential adsorption sites on graphene. Notably, the effect of carbon atom count on adsorption is more pronounced for graphene-PAMAM than for CNT-PAMAM.

\begin{figure*}[!h]
     \centering
    \includegraphics[width=1\linewidth]{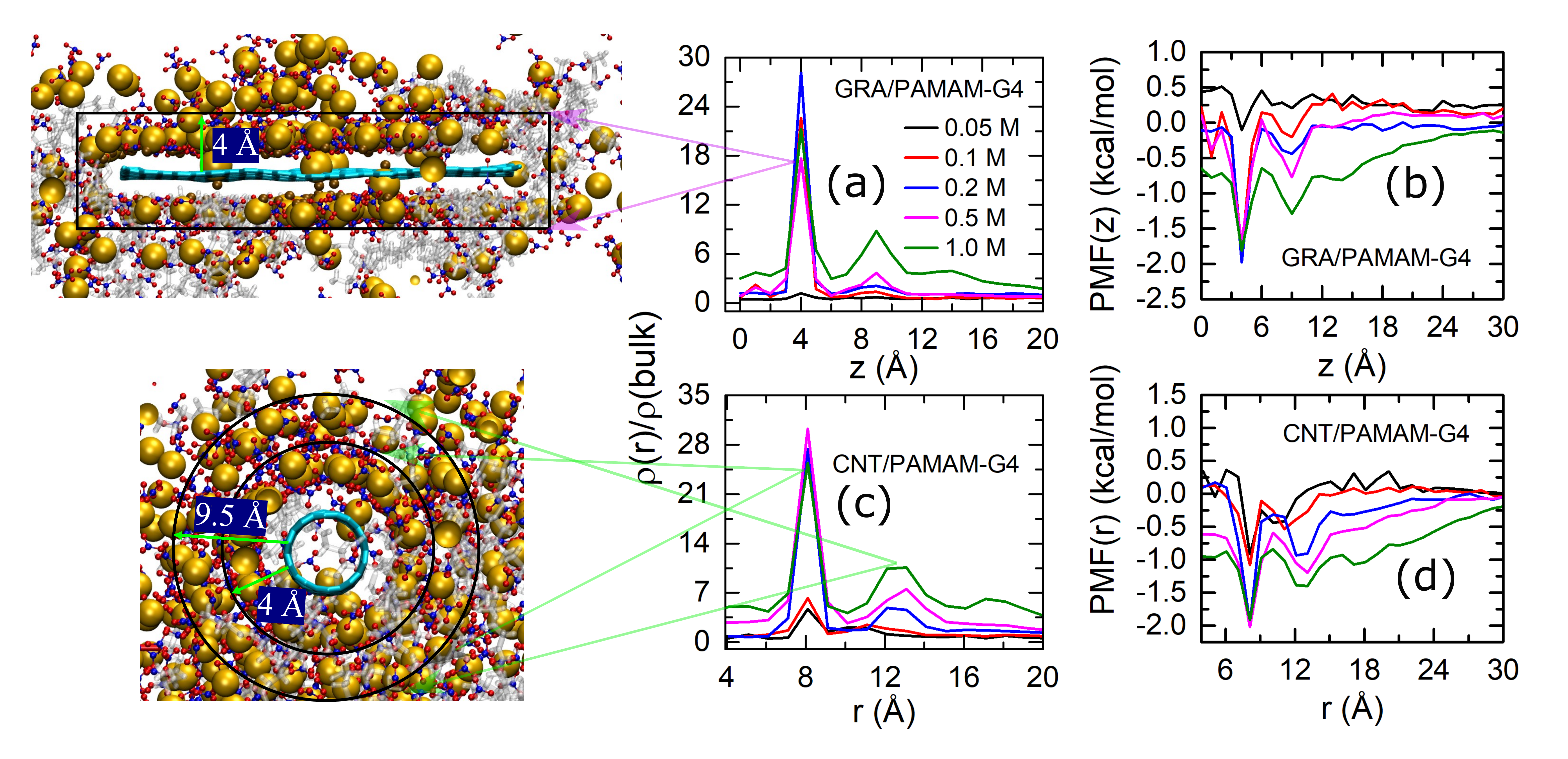}
    \caption[Normalized radial density profile of uranyl (\ce{UO2^2+}) ions normal ($\perp$) to the surface of (a) GRA and (c) CNT surfaces]{Normalized radial density profile of uranyl (\ce{UO2^2+}) ions normal ($\perp$) to the surface of (a) GRA and (c) CNT surfaces, respectively at low pH and various uranyl concentrations,
 shown alongside representative snapshots of adsorption. The black line in the snapshots demarcates the first solvation layer boundary. Corresponding potential of mean force (PMF) profiles for uranyl ion adsorption are presented in (b) and (d), respectively.}
    \label{fig:normal_density_plot_combined}   
\end{figure*}

%\subsection{Normal Density Profiles}
%\textcolor{blue}{
To understand the spatial distribution of adsorbed uranyl ions, we calculated the average normalized number density profiles ($\rho_{\perp}$) normal to the graphene/CNT surface (z-axis of graphene and the radial axis of CNT) for various uranyl ion concentrations (Figure. \ref{fig:normal_density_plot_combined}).
A prominent peak at \SI{4}{\angstrom} from the graphene/CNT surface indicates uranyl ions also adsorbed onto the graphene/CNT surface, likely due to the hydrophobic nature of the graphene/CNTs and their high specific surface area. Interestingly, this peak position remains constant with increasing uranyl concentration. The peak intensity initially increases from 0.05 M to 0.5 M, suggesting monolayer formation. However, it decreases at 1 M due to uranyl ion saturation and overcrowding at the graphene/CNT interface by the cationic uranyl ions.
Additionally, a second peak emerges at \SI{9.5}{\angstrom} with increasing concentrations, signifying counterion layering at higher uranyl ion concentrations. The second peak is more prominent for the CNT-PAMAM composite case compared to the graphene-PAMAM cases. 
\begin{figure*}[!h]
    \centering
    \includegraphics[width=0.75\linewidth]{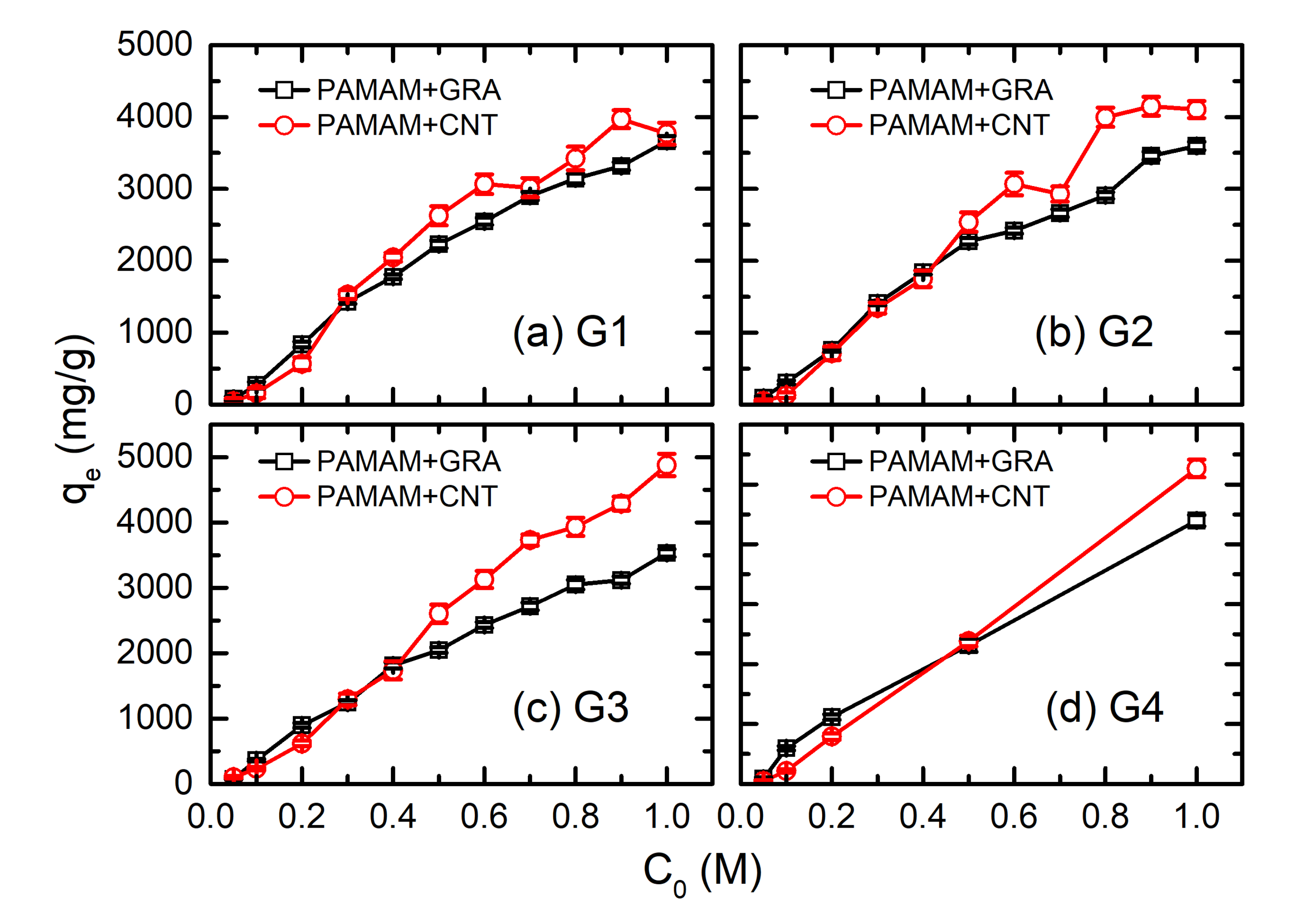}
    \caption[Uranyl ion (\ce{UO2^2+}) adsorption capacity ($q_e$) on PAMAM-functionalized graphene-PAMAM nanocomposite (PAMAM+GRA) and CNT-PAMAM (PAMAM+CNT) nanocomposites]{Uranyl ion (\ce{UO2^2+}) adsorption capacity ($q_e$) on PAMAM-functionalized graphene-PAMAM nanocomposite (PAMAM+GRA) and CNT-PAMAM (PAMAM+CNT) nanocomposites as a function of initial uranyl salt concentration ($C_0$) for different PAMAM dendrimer generations: (a) G1, (b) G2, (c) G3, and (d) G4. at low pH of the simulation.}
    \label{fig:qe}
\end{figure*}
Understanding the strength and location of uranyl ion adsorption is crucial for optimizing the design of efficient uranium removal materials. Here, we calculate potential of mean force (PMF) to quantify the free energy associated with uranyl ion adsorption onto graphene/CNT-PAMAM nanocomposites.
PMF profiles were calculated for the graphene/CNT-PAMAM system at various uranyl ion concentrations using the following equation:
\begin{equation}
    \text{PMF (z/r)} = -K_\text{B} T \ln\left(\frac{\rho_{\perp}(z/r)}{\rho_{\text{bulk}}}\right)
\end{equation}
where $\rho_{\perp}(z/r)$ represents the number density of uranyl ions normal to the graphene/CNT plane (z-axis for graphene or radial axis for CNT) obtained from density profile analysis (shown in Figure. \ref{fig:normal_density_plot_combined} (a,c)), and $\rho_{\text{bulk}}$ denotes the number density of uranyl ions at the bulk region. Figure \ref{fig:normal_density_plot_combined} (b,d) shows the PMF profiles depicting the uranyl adsorption of the graphene/CNT-PAMAM adsorbents.

Interestingly, the position of PMF minima (indicating the most favorable adsorption sites) and the peak position in the corresponding density profiles (Figure. \ref{fig:normal_density_plot_combined} (a,b)) remain same across different uranyl ion concentrations. This suggests the preference for a specific, energetically favorable nanocomposite site. This distance is consistent with the number density plot, which also signifies uranyl ion adsorption on graphene and CNT surfaces.  

Figures \ref{fig:normal_density_plot_combined} (b,d) depict that the PMF profile progressively deepens (becomes more negative) from $-0.5$ to \SI{-2.0}{} kcal/mol with uranyl ion concentration up to 0.5 M and a slight variation is observed at concentrations exceeding 0.5 M. This trend signifies a saturation of the adsorption strength at higher concentrations, likely due to the saturation of available binding sites on the nanocomposite. 
%Weakening of adsorption strength at higher concentrations (above 0.5 M) was likely due to the saturation of binding sites. 
The deepening PMF profile with increasing concentration (up to 0.5 M) confirmed a stronger free energy gain for adsorption, highlighting the optimal concentration range for efficient uranyl removal. 

%\section{Adsorption Capacity:}
The potential of graphene/CNT-PAMAM nano-composites for removing uranyl ions from aqueous solutions is also quantified by estimating the adsorption capacity  of the nanocomposite. The adsorption capacity has been quantified by calculating the number of uranyl ions adsorbed per unit mass of the nanocomposite at various uranyl ion concentrations using the following equation:
\cite{xiong2019hypervalent,zhang2015adsorption,singha2015adsorption}
\begin{equation}
    q_e = \frac{\text{Mass of adsorbed Uranyl ions}}{\text{Mass of Adsorbent}}
\end{equation}
Adsorption capacity is plotted as a function of Uranyl concentration for different generations of PAMAM dendrimer for both the GRA-PAMAM and CNT-PAMAM nanocomposite in Figure \ref{fig:qe}.
%We observed the graphene-PAMAM nanocomposite exhibits a higher number of total uranyl ion adsorption compared to the CNT-PAMAM nanocomposite as shown in Figure \ref{fig:num_U_ads_combined} (f). However, 
Figure \ref{fig:qe} reveals that CNT-PAMAM possesses greater adsorption capacity $q_e$ than the graphene-PAMAM nanocomposite for the uranyl ion concentration  0.05 M  to  1 M. This is primarily due to the curvature of the CNT, which caused the PAMAM to cover its surface over the dynamics due to higher hydrophobic intermolecular interactions. Therefore, its adsorption/binding sites were more accessible for uranyl ions binding. 

\begin{figure*}[h!]
    \centering
    \includegraphics[width=1\linewidth]{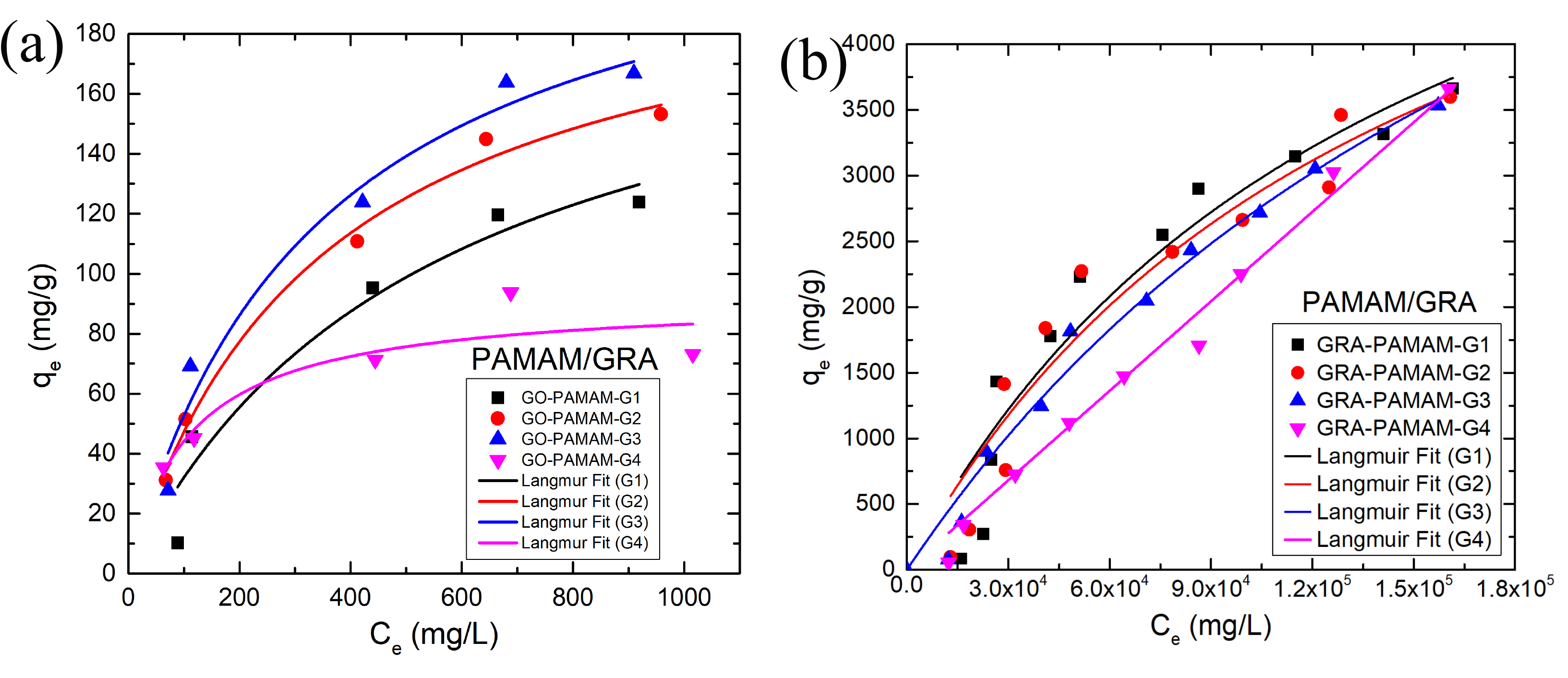}
    \caption[adsorption isotherms of adsorption of uranyl ions for graphene-PAMAM nanocomposite]{(a) Experimental and (b) MD simulated uranyl (\ce{UO_2^{2+}}) ion adsorption isotherms for different generations of PAMAM grafted onto graphene-Oxide (GO-PAMAM) in the experiment and graphene-PAMAM nanocomposite in MD simulation at low pH. The lines represent the Langmuir isotherm model fit to the data.}
    \label{fig:gra-Lang}
\end{figure*}
% \begin{table*}[h!]
% \small
% \centering
% \caption{Langmuir adsorption isotherm fitting parameters for GRA/PAMAM at low pH.}
% \begin{tabular}{|c|c|c|c|c|}
% \hline
% Generation & \multicolumn{2}{c|}{\textbf{$q_{\text{max}}$}} & \multicolumn{2}{c|}{\textbf{$K_L$}} \\ \cline{2-5}
% & Expt. & MD simulation & Expt. & MD Simulation \\ \hline
% G1 & 200.17 & 7090 $\pm$ 1830 & 0.017 & 6.92E-06 $\pm$ 3.01E-06 \\ \hline
% G2 & 212.78 & 6900.69798 $\pm$ 1796.83958 & 0.0027 & 6.85E-06 $\pm$ 3.02E-06 \\ \hline
% G3 & 229.30 & 8716.07099 $\pm$ 1859.46627 & 0.0029 & 4.42E-06 $\pm$ 1.37E-06 \\ \hline
% G4 & 92.89 & 1.97E+09 $\pm$ 1.12E+014 & 0.0086 & 1.15E-11 $\pm$ 6.57E-07 \\ \hline
% \end{tabular}
%\label{tab:Lang-GRA}
%\end{table*}
\begin{table*}[h!]
\small
\centering
\caption{Langmuir Adsorption Isotherm Fitting Parameters for graphene-PAMAM at low pH.}
\begin{tabular}{|c|c|c|c|c|}
\hline
Generation & \multicolumn{2}{c|}{$\boldsymbol{q_{\text{max}}}$ (mg/g)} & \multicolumn{2}{c|}{$\boldsymbol{K_L}$ (L/mg)} \\ \cline{2-5}
& Expt. & MD Simulation & Expt. & MD Simulation \\ \hline
G1 & 200 & $7.09 \times 10^3 \pm 1.83 \times 10^3$  & 1.7 $\times 10^{-2}$ & $6.92 \times 10^{-6} \pm 3.01 \times 10^{-6}$  \\ \hline
G2 & 213 & $6.90 \times 10^3 \pm 1.80 \times 10^3$ & 2.7 $\times 10^{-3}$ & $6.85 \times 10^{-6} \pm 3.02 \times 10^{-6}$  \\ \hline
G3 & 229 & $8.72 \times 10^3 \pm 1.86 \times 10^3$ & 2.9 $\times 10^{-3}$ & $4.42 \times 10^{-6} \pm 1.37 \times 10^{-6}$  \\ \hline
G4 & 92.9 & $1.97 \times 10^9 \pm 1.12 \times 10^{14}$ & 8.6 $\times 10^{-3}$  & $1.15 \times 10^{-11} \pm 6.57 \times 10^{-7}$ \\ \hline
\end{tabular}
\label{tab:Lang-GRA}
\end{table*}

\begin{figure*}[h!]
    \centering
    \includegraphics[width=\linewidth]{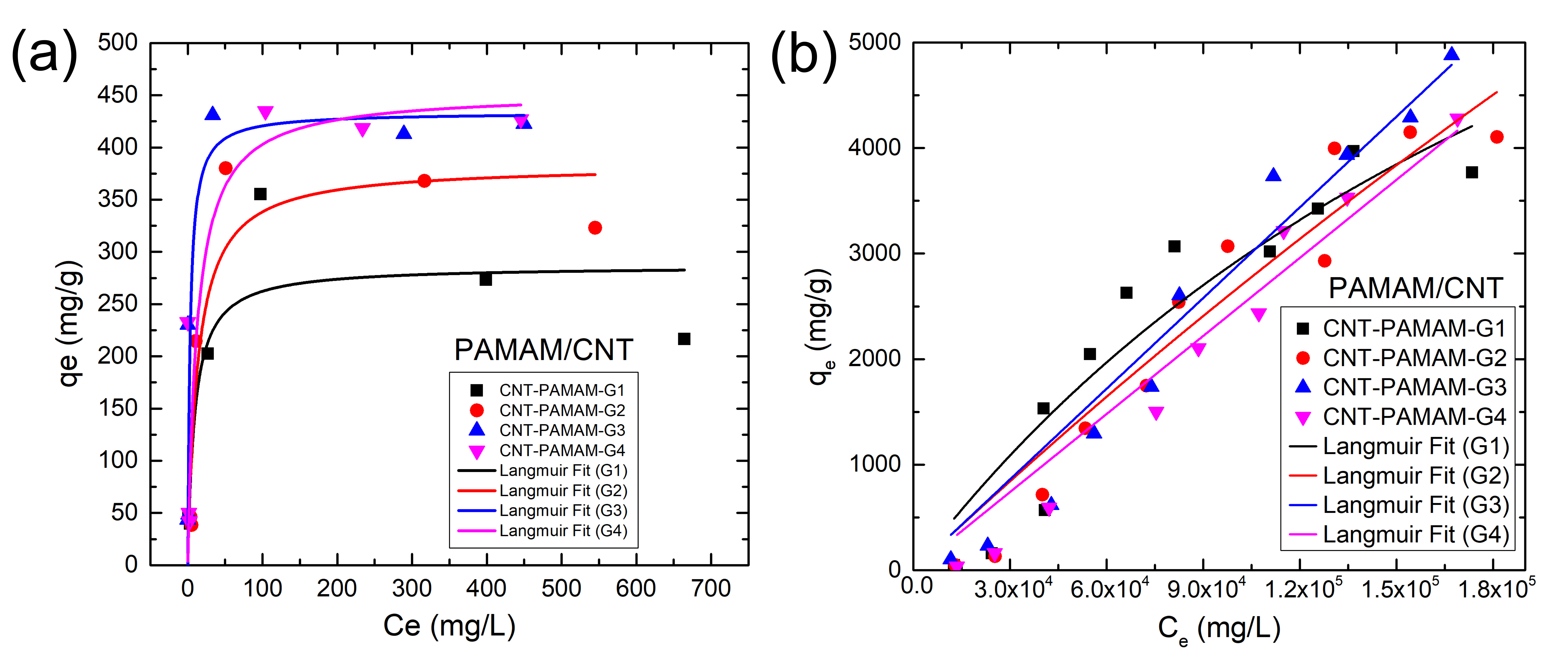}
    \caption[adsorption isotherms of adsorption of uranyl ions for CNT-PAMAM nanocomposite]{(a) Experimental and (b) MD simulated uranyl (\ce{UO_2^{2+}}) ion adsorption isotherms for different generations of PAMAM grafted onto multiwalled CNT in the experiment and single-walled CNT-PAMAM nanocomposite in MD simulation at low pH. The lines represent the Langmuir isotherm model fit to the data at low pH .}
    \label{fig:CNT-Lang}
\end{figure*}
\begin{table*}[h!]
\small
\centering
\caption{Langmuir Adsorption Isotherm Fitting parameters for CNT-PAMAM at low pH.}
\begin{tabular}{|c|c|c|c|c|}
\hline
Generation & \multicolumn{2}{c|}{$\boldsymbol{q_{\text{max}}}$ (mg/g)} & \multicolumn{2}{c|}{$\boldsymbol{K_L}$ (L/mg)} \\ \cline{2-5}
& Expt. & MD Simulation & Expt. & MD Simulation \\ \hline
G1 & 290 & $1.06 \times 10^4 \pm 5.48 \times 10^3$  & 1.2 $\times 10^{-1}$ & $3.82 \times 10^{-6} \pm 2.84 \times 10^{-6}$  \\ \hline
G2 & 376 & $3.49 \times 10^4 \pm 5.77 \times 10^4$ & 9.0 $\times 10^{-2}$ & $8.23 \times 10^{-7} \pm 1.51 \times 10^{-6}$  \\ \hline
G3 & 451 & $3.56 \times 10^9 \pm 5.32 \times 10^{14}$ & 7.0 $\times 10^{-2}$ & $8.04 \times 10^{-12} \pm 1.20 \times 10^{-6}$  \\ \hline
G4 & 456 & $3.70 \times 10^9 \pm 7.69 \times 10^{14}$ & 7.0 $\times 10^{-2}$  & $6.66 \times 10^{-12} \pm 1.38 \times 10^{-6}$ \\ \hline
\end{tabular}
\label{tab:CNT-lang}
\end{table*}

%\subsection{Adsorption Isotherm}
To better understand the adsorption behaviors of uranyl ions on the nanocomposite and analyze the adsorption capacities of the adsorbents, we have studied the adsorption isotherms using Langmuir models, described by the equation:
\begin{equation}\label{eq:q_e}
    q_e = q_{\text{max}} \frac{ K_{\text{L}} C_e}{1 + K_{\text{L}}  C_e}
\end{equation}
where \( q_e \) (\SI{}{\milli\gram\per\gram}) represents the amount of uranyl ions adsorbed per unit mass of the adsorbent (\textit{i.e.} graphene/CNT-PAMAM nanocomposite). \( q_{\text{max}} \) (\SI{}{\milli\gram\per\gram}) is the maximum adsorption capacity, 
\( K_{\text{L}} \) (\SI{}{\liter\per\milli\gram}) is the Langmuir isotherm constant, and \( C_e \) (\SI{}{\milli\gram\per\liter}) denotes the equilibrium uranyl ion concentration. 

Figure \ref{fig:gra-Lang} depicts the maximum adsorption capacity ($q_{max}$)  increases with increasing dendrimer generation for both experiment and MD simulation (Table \ref{tab:Lang-GRA}) for the graphene-PAMAM nanocomposite. This indicates that higher generations of PAMAM dendrimers can adsorb more uranyl ions due to higher binding sites. However, there is a decrease in $q_{max}$ for G4 in the experimental data, which could be attributed to steric hindrance or other factors that limit the accessibility of binding sites in the larger dendrimer.
%Although the experiment (multilayered graphene oxide) and simulation (pristine graphene) systems are quite different, they have shown similar trends. 
Table \ref{tab:Lang-GRA} also illustrates that $K_L$ decreases with increasing uranyl ion concentrations. This decrease in binding affinity is likely due to electrostatic repulsion between the adsorbed uranyl ions, which becomes more significant at higher concentrations. Note that a quantitative comparison of experimental and simulated adsorption behavior is difficult as the experiments use multilayered graphene oxide (GO)  instead of pristine graphene as used in the simulation. Another notable difference is the large uranyl concentration regime simulated compared to the lower uranyl concentration achieved in the experiment.

Figure \ref{fig:CNT-Lang} and Table \ref{tab:CNT-lang} suggest that Langmuir isotherm does not capture the adsorption behavior for CNT-PAMAM nanocomposites very well. The deviation from the Langmuir model suggests that the adsorption process may not be monolayer, as assumed by the model. This is supported by the density profiles in Figure \ref{fig:normal_density_plot_combined}, which show a second peak at higher concentrations, indicating the formation of additional layers of uranyl ions beyond the first solvation shell. This multilayer adsorption could be due to the curvature of CNTs, which may create additional binding sites or promote the formation of uranyl ion clusters.

\section{Conclusions}{\label{sec:level4}}
In this work, we employed a combined experimental and molecular dynamics (MD) simulation approach to elucidate the adsorption kinetics of uranyl ions onto graphene/CNT-PAMAM nanocomposites. Our findings demonstrate that adsorption is influenced by several key factors, including solution pH, dendrimer generation, and initial uranium concentration. Notably, CNT-PAMAM nanocomposites exhibited superior adsorption capacity compared to their graphene counterparts, likely due to enhanced accessibility of binding sites facilitated by the curvature of CNTs. MD simulations revealed that uranyl ions adsorb onto both the PAMAM surface and the underlying graphene/CNT surface. The Langmuir isotherm model effectively described the adsorption behavior of graphene-PAMAM nanocomposites, with adsorption capacity increasing as a function of dendrimer generation. However, this model did not accurately capture the multilayer adsorption observed at higher uranyl ion concentrations in CNT-PAMAM nanocomposites, highlighting the need for more nuanced models to describe the adsorption behavior in these systems. Overall, this study provides valuable insights into the molecular mechanisms underlying uranyl ion adsorption onto graphene/CNT-PAMAM nanocomposites. The findings suggest that these materials hold promise for efficient uranium removal from aqueous solutions, and further optimization of their design could lead to the development of highly effective adsorbents for environmental remediation applications.
%%%%%%%%%%%%%%%%%%%%%%%%%%%%%%%
%%Author Information
%%%%%%%%%%%%%%%%%%%%%%%%%%%%%%%
\section{Author Information}
\subsection{Corresponding Author}

\textbf{Prabal Kumar Maiti} - Center for Condensed Matter Theory, Department of Physics, Indian Institute of Science Bangalore, India, 560012; Orcid: 
http://orcid.org/0000-0002-9956-1136;
Email: maiti@iisc.ac.in

\subsection{Authors}
\textbf{Tarun Maity} - Center for Condensed Matter Theory, Department of Physics, Indian Institute of Science, Bangalore 560012, India;  Orcid:  https://orcid.org/0000-0002-4405-0371; Email: tarunmaity@iisc.ac.in

\textbf{Yogendra Kumar} - Center for Condensed Matter Theory, Department of Physics, Indian Institute of Science Bangalore, India, 560012;
Orcid: https://orcid.org/0000-0003-1487-7017; Email: kyogendra@iisc.ac.in

\textbf{Ashish Kumar Singha Deb} - Chemical Engineering Division, Bhabha Atomic Research Centre, Mumbai 400 085, India;  Orcidhttps://orcid.org/0000-0003-2283-5810; Email: aksdeb@barc.gov.in

\textbf{Sheikh Musharaf Ali} - Chemical Engineering Division, Bhabha Atomic Research Centre, Mumbai 400 085, India;  Homi Bhabha National Institute, Anushaktinagar, Mumbai 40085, India;  Orcid: https://orcid.org/0000-0003-0457-0580; Email: musharaf@barc.gov.in

%% classes. This should be given within the "acknowledgement"
\section{Acknowledgement}
We acknowledge funding through BRNS (58/14/09/2021-BRNS/37116).
We also acknowledge the computational support provided by the DST-funded TUE-CMS program at IISc.
% \begin{acknowledgement}
% We acknowledge funding through BRNS(58/14/09/2021-BRNS/37116).
% We also acknowledge the computational support provided by the DST-funded TUE-CMS program at IISc. 
% \end{acknowledgement}

%\begin{suppinfo}

%\end{suppinfo}


% \newpage
\bibliography{ref}
\end{document}

\maketitle

\newpage

\renewcommand{\thefigure}{S\arabic{figure}}
\renewcommand{\thetable}{S\arabic{table}}

\begin{figure}
    \centering
    \includegraphics[width=\linewidth]{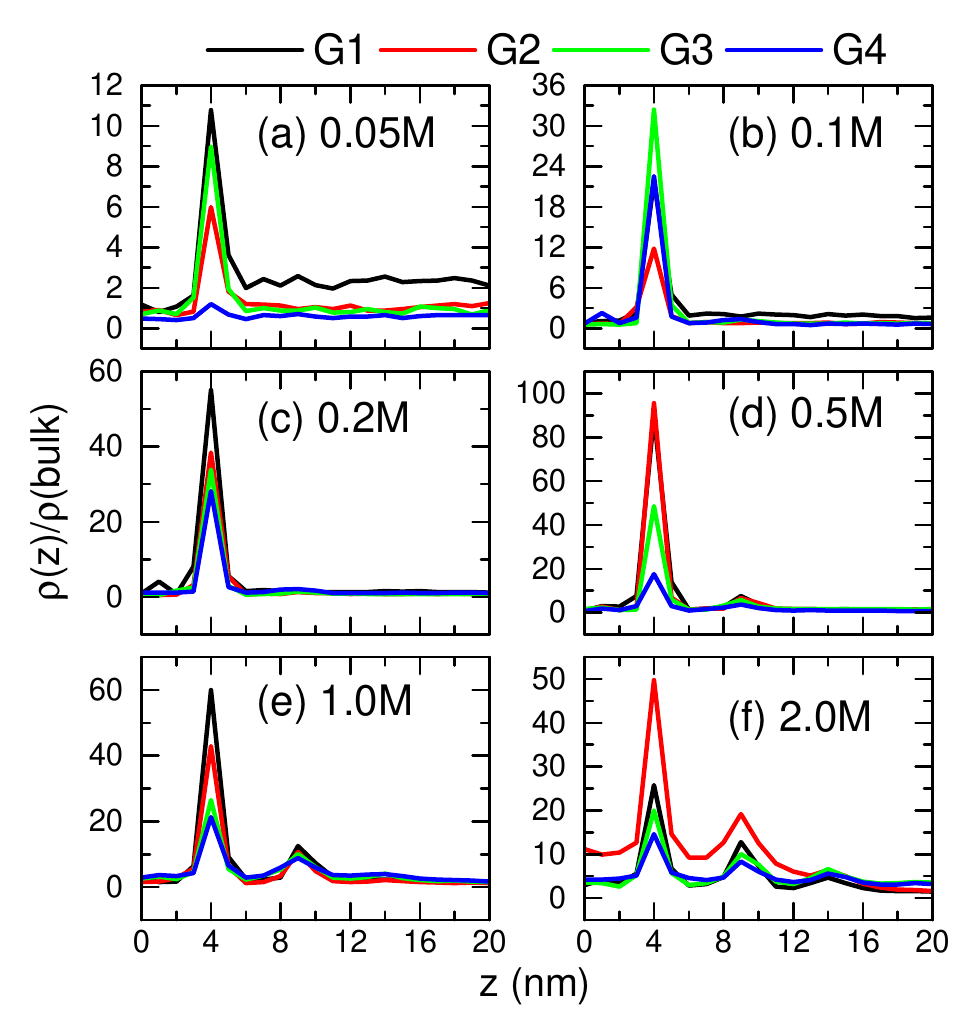}
    \caption{Normalized radial density profile of uranyl (\ce{UO2^2+}) ions normal ($\perp$) to the surface of  GRA for various uranyl ion concentration}
    \label{fig:enter-label}
\end{figure}

\begin{table*}[!ht]
\setlength{\tabcolsep}{3pt}
\small
    \centering
    \begin{tabular}
    [\linewidth]{|p{1.9 cm}|p{1.5 cm}|c|c|c|c|c|c|c|c|}
    \hline
        \textbf{Generation} & \textbf{Uranium Conc. (M)} & \multicolumn{4}{c|}{\textbf{Graphene/PAMAM}}   &  \multicolumn{4}{c|}{\textbf{CNT/PAMAM}}   \\ \hline
      %   & \multicolumn{8}{c|}{Generation 1 (G1)} \\ \hline
        & & \multicolumn{2}{c|}{\textbf{Low pH}} & \multicolumn{2}{c|}{\textbf{Neutral pH}} & \multicolumn{2}{c|}{\textbf{Low pH}} & \multicolumn{2}{c|}{\textbf{Neutral pH}}  \\ \hline
        & ~ &\textbf{ Uranium} & \textbf{Nitrate} & \textbf{Uranium} & \textbf{Nitrate} & \textbf{Uranium} & Nitrate &\textbf{ Uranium} & \textbf{Nitrate}  \\ \hline
       & 0.05 & 45 & 180 & 36 & 128 & 33 & 148 & 30 & 112 \\ \cline{2-10}
        &0.1 & 90 & 300 & 75 & 232 & 69 & 244 & 63 & 200  \\ \cline{2-10}
      \multirow{2}{*}{\textbf{G1}} & 0.2 & 180 & 540 & 153 & 440 & 138 & 428 & 129 & 376  \\ \cline{2-10}
       & 0.5 & 456 & 1276 & 387 & 1064 & 348 & 988 & 324 & 896 \\ \cline{2-10}
       & 1 & 912 & 2492 & 777 & 2104 & 699 & 1924 & 651 & 1768 \\ \cline{2-10}
        \hline
        % G2 & ~ & ~ & ~ & ~ & ~ & ~ & ~ & ~ & ~ & ~ \\ \hline
       & 0.05 & 54 & 286 & 57 & 216 & 54 & 268 & 57 & 216 \\ \cline{2-10}
       & 0.1 & 108 & 412 & 114 & 368 & 108 & 412 & 114 & 368  \\ \cline{2-10}
      \multirow{2}{*}{\textbf{G2}} & 0.2 & 219 & 708 & 228 & 672 & 219 & 708 & 228 & 672 \\ \cline{2-10}
       & 0.5 & 549 & 1588 & 573 & 1592 & 552 & 1596 & 573 & 1592 \\ \cline{2-10}
       & 1 & 1101 & 3060 & 1146 & 3120 & 1107 & 3076 & 1146 & 3120 \\ \hline
       & 0.05 & 75 & 452 & 78 & 336 & 78 & 460 & 75 & 328 \\ \cline{2-10}
       & 0.1 & 153 & 660 & 156 & 544 & 159 & 676 & 153 & 536  \\ \cline{2-10}
      \multirow{2}{*}{\textbf{G3}} & 0.2 & 306 & 1068 & 312 & 960 & 321 & 1108 & 309 & 952  \\ \cline{2-10}
       & 0.5 & 765 & 2292 & 780 & 2208 & 807 & 2404 & 774 & 2192 \\ \cline{2-10}
        &1 & 1533 & 4340 & 1563 & 4296 & 1614 & 4556 & 1548 & 4256 \\ \hline
        %  \multicolumn{9}{|c|}{Generation 4 (G4)} \\ \hline
        % G4 & ~ & ~ & ~ & ~ & ~ & ~ & ~ & ~  \\ \hline
        &0.05 & 114 & 812 & 102 & 528 & 114 & 812 & 108 & 544  \\ \cline{2-10}
        &0.1 & 231 & 1124 & 207 & 808 & 228 & 1116 & 216 & 832  \\ \cline{2-10}
       \multirow{2}{*}{\textbf{G4}} &0.2 & 465 & 1748 & 414 & 1360 & 459 & 1732 & 435 & 1416  \\ \cline{2-10}
        &0.5 & 1164 & 3612 & 1038 & 3024 & 1152 & 3580 & 1092 & 3168 \\ \cline{2-10}
        &1 & 2331 & 6724 & 2076 & 5792 & 2304 & 6652 & 2184 & 6080  \\ \hline
    \end{tabular}
        \caption{Details of the graphene/PAMAM and CNT/PAMAM simulated systems.}
    \label{tab:systemDetails}
\end{table*}

\begin{table}[!ht]
\small
    \centering
    \begin{tabular}[\linewidth]{|p{1.8 cm}|p{1.5 cm}|c|c|c|c|c|c|c|c|}
    \hline
        Generation & Uranium Conc. (M) & \multicolumn{4}{c|}{Graphene/PAMAM}  &  \multicolumn{4}{c|}{CNT/PAMAM}   \\ \hline
      %   & \multicolumn{8}{c|}{Generation 1 (G1)} \\ \hline
        & & \multicolumn{2}{c|}{Low pH} & \multicolumn{2}{c|}{Neutral pH} & \multicolumn{2}{c|}{Low pH} & \multicolumn{2}{c|}{Neutral pH}  \\ \hline
        & ~ & Uranium & Nitrate & Uranium & Nitrate & Uranium & Nitrate & Uranium & Nitrate  \\ \hline
       & 0.05 & 45 & 180 & 36 & 128 & 33 & 148 & 30 & 112 \\ \cline{2-10}
        &0.1 & 90 & 300 & 75 & 232 & 69 & 244 & 63 & 200  \\ \cline{2-10}
      \multirow{2}{*}{G1} & 0.2 & 180 & 540 & 153 & 440 & 138 & 428 & 129 & 376  \\ \cline{2-10}
       & 0.5 & 456 & 1276 & 387 & 1064 & 348 & 988 & 324 & 896 \\ \cline{2-10}
       & 1 & 912 & 2492 & 777 & 2104 & 699 & 1924 & 651 & 1768 \\ \cline{2-10}
        \hline
        % G2 & ~ & ~ & ~ & ~ & ~ & ~ & ~ & ~ & ~ & ~ \\ \hline
       & 0.05 & 54 & 286 & 57 & 216 & 54 & 268 & 57 & 216 \\ \cline{2-10}
       & 0.1 & 108 & 412 & 114 & 368 & 108 & 412 & 114 & 368  \\ \cline{2-10}
      \multirow{2}{*}{G2} & 0.2 & 219 & 708 & 228 & 672 & 219 & 708 & 228 & 672 \\ \cline{2-10}
       & 0.5 & 549 & 1588 & 573 & 1592 & 552 & 1596 & 573 & 1592 \\ \cline{2-10}
       & 1 & 1101 & 3060 & 1146 & 3120 & 1107 & 3076 & 1146 & 3120 \\ \hline
       & 0.05 & 75 & 452 & 78 & 336 & 78 & 460 & 75 & 328 \\ \cline{2-10}
       & 0.1 & 153 & 660 & 156 & 544 & 159 & 676 & 153 & 536  \\ \cline{2-10}
      \multirow{2}{*}{G3} & 0.2 & 306 & 1068 & 312 & 960 & 321 & 1108 & 309 & 952  \\ \cline{2-10}
       & 0.5 & 765 & 2292 & 780 & 2208 & 807 & 2404 & 774 & 2192 \\ \cline{2-10}
        &1 & 1533 & 4340 & 1563 & 4296 & 1614 & 4556 & 1548 & 4256 \\ \hline
        %  \multicolumn{9}{|c|}{Generation 4 (G4)} \\ \hline
        % G4 & ~ & ~ & ~ & ~ & ~ & ~ & ~ & ~  \\ \hline
        &0.05 & 114 & 812 & 102 & 528 & 114 & 812 & 108 & 544  \\ \cline{2-10}
        &0.1 & 231 & 1124 & 207 & 808 & 228 & 1116 & 216 & 832  \\ \cline{2-10}
       \multirow{2}{*}{G4} &0.2 & 465 & 1748 & 414 & 1360 & 459 & 1732 & 435 & 1416  \\ \cline{2-10}
        &0.5 & 1164 & 3612 & 1038 & 3024 & 1152 & 3580 & 1092 & 3168 \\ \cline{2-10}
        &1 & 2331 & 6724 & 2076 & 5792 & 2304 & 6652 & 2184 & 6080  \\ \hline
    \end{tabular}
\end{table}

\subsection{Materials and reagents}

Concentrated nitric acid (\ce{HNO3}) and sulfuric acid (\ce{H2SO4}) were purchased from Thomas Baker (Chemicals) Pvt. Limited. Ethylenediamine (EDA), methyl acrylate (MA), tetrahydrofuran (THF) and thionyl chloride (\ce{SOCl_2}), methanol (\ce{MeOH}) and ethanol (\ce{EtOH}) were purchased from S.D. Fine Chem Ltd. and used without further purification.

\subsection{Synthesis of CNT-PAMAM}

\begin{figure}
    \centering
    \includegraphics[width=1\linewidth]{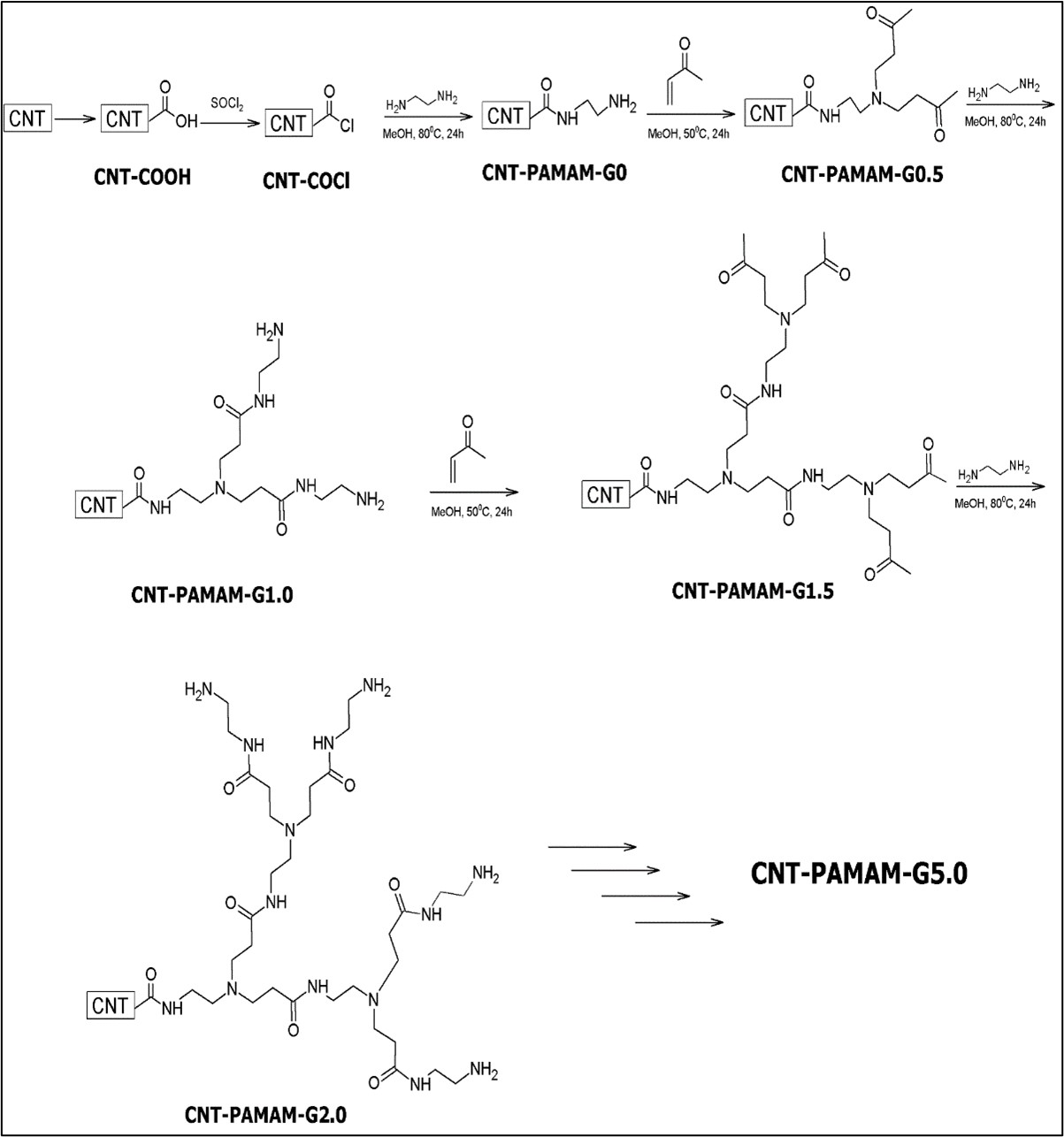}
    \caption{Reaction sequence of the de-novo synthesis of CNT-PAMAM-G5 from CNT}
    \label{fig:expt-CNT}
\end{figure}

Multi-walled carbon nanotubes (MWCNTs), synthesized through catalytic CVD method as reported earlier\cite{deb2017novel}, were used and referred to as carbon nanotubes (CNTs). PAMAM dendrons starting from generation zero (G0) to generation five (G5) were grown on the surface of MWCNTs through the ‘grafting from’ method by a divergent synthesis as described below. The reaction sequence of the de-novo synthesis of CNT-PAMAM-G5 from MWCNT is depicted in Figure \ref{fig:expt-CNT}.
\subsection{Preparation of CNT-COOH from CNT}
\SI{1}{\gram} purified MWCNTs were mixed with 200 mL of 3:1 (v/v) mixture of concentrated \ce{H2SO4} and \ce{HNO3}. This mixture was sonicated in an ultrasonic bath for 6 hours at \SI{60}{\celsius}. The black syrupy suspension was then cooled, diluted five times with distilled water, filtered under vacuum, and washed with distilled water until the pH of the filtrate became neutral. The filtered blackish product was dried at \SI{110}{\celsius} under vacuum overnight.
\subsection{Generation of \ce{CNT-COCl} from \ce{CNT-COOH}}
CNT-COOH (5.309g) was suspended in \ce{SOCl2} (300ml) for 30 min 
and stirred for 24 hours at \SI{70}{\celsius}. The solution 
was filtered, washed initially with copious amounts of anhydrous THF to remove \ce{SOCl2}, 
and then with methanol. Finally, the product was dried at \SI{80}{\celsius} under vacuum overnight, generating \SI{5.333}{\gram} \ce{CNT-COCl}.

\subsection{Synthesis of \ce{CNT-NH2} (CNT-PAMAM-G0) initiator from \ce{CNT-COCl}}
CNT-COCl (5.00g) was mixed with EDA (50ml) and sonicated for 30 min at room temperature (RT). The mixture was stirred for another 24h at \SI{60}{\celsius}. The mixture was then transferred in an oak ridge centrifuge tube with \ce{EtOH} and centrifuged at 10000 rpm for 15 minutes. It was then allowed to stand for 10 minutes. The supernatant solution was removed, and the black residue obtained was washed three times with \ce{EtOH} and once with \ce{MeOH}. Each time, the colorless supernatant solution was decanted. Finally, the product was transferred to a glass petri-dish using \ce{MeOH}, and kept overnight for drying at \SI{80}{\celsius} in a vacuum oven. The dried product thus obtained (\ce{CNT-NH2}) was collected aside and weighed as \SI{5.26}{\gram}.

\subsection{Synthesis of CNT-PAMAM-G0.5 from CNT-PAMAM-G0 (Michael addition)}
CNT-PAMAM-G-0 (5.00g) was suspended in a mixture of 150 mL of MeOH and 50 mL of MA through sonication for 30 min at RT. The resulting suspension was stirred for 48 hours at 60°C. The mixture was cooled and transferred in oak ridge centrifuge tubes with \ce{EtOH}, and centrifuged at 10000 rpm for 15 minutes. It was then allowed to stand for 10 minutes. The supernatant solution was removed and the black residue obtained was washed three times with EtOH as above using centrifugation and one time with MeOH. Each time, the colored or colorless supernatant solution was decanted. Finally, the product was transferred to a glass petri-dish using \ce{MeOH}, and kept overnight for drying at 80oC in a vacuum oven.  The dried product thus obtained (CNT-PAMAM-G0.5) was collected aside and weighed as 4.52 g.

\subsection{Synthesis of CNT-PAMAM-G1.0 from CNT-PAMAM-G0.5 (ester amidation)}
4.00 g of MWCNT G-0.5 was suspended in a mixture of 150 mL MeOH and 50 mL of EDA through sonication for 30 min at RT. The mixture was stirred for another 48h at 60°C. The mixture was cooled and transferred in oak ridge centrifuge tubes with \ce{EtOH}, and centrifuged at 10000 rpm for 15 minutes. It was then allowed to stand for 10 minutes. The supernatant solution was removed, and the black residue obtained was washed three times with EtOH as above using centrifugation and one time with MeOH. Each time, the colored or colorless supernatant solution was decanted. Finally, the product was transferred to a glass petri-dish using \ce{MeOH}, and kept overnight for drying at \SI{80}{\celsius} in a vacuum oven.  The dried product thus obtained (CNT-PAMAM-G1.0) was collected aside and weighed as \SI{4.08}{\gram}. 

\subsection{Synthesis of CNT-PAMAM-G1.5 from CNT-PAMAM-G1.0 (Michael addition)}

CNT-PAMAM-G1.0 (\SI{3.70}{\gram}) was suspended in a mixture of \SI{150}{\milli\liter} of \ce{MeOH} and \SI{50}{\milli\liter} of \ce{MA} through sonication for \SI{30}{\minute} at RT. The resulted suspension was stirred for \SI{48}{\hour} at \SI{60}{\degreeCelsius}. The mixture was cooled and transferred in oak ridge centrifuge tubes with \ce{EtOH} and centrifuged at \SI{10000}{rpm} for \SI{15}{\minute}. The supernatant solution was removed and the black residue obtained was washed three times with \ce{EtOH} as above using centrifugation and one time with \ce{MeOH}. Each time, the colored or colorless supernatant solution was decanted. Finally, the product was transferred to a glass petri-dish using \ce{MeOH}, and kept overnight for drying at \SI{80}{\degreeCelsius} in a vacuum oven. The dried product thus obtained (CNT-PAMAM-G1.5) was collected and weighed.

\subsection{Synthesis of MWCNT-PAMAM-G2.0 from MWCNT-PAMAM-G1.5 (ester amidation)}

\SI{3.50}{\gram} of CNT-PAMAM-G1.5 was suspended in a mixture of \SI{150}{\milli\liter} \ce{MeOH} and \SI{50}{\milli\liter} of \ce{EDA} through sonication for \SI{30}{\minute} at RT. The mixture was stirred for another \SI{48}{\hour} at \SI{60}{\degreeCelsius}. The mixture was cooled and transferred in oak ridge centrifuge tubes with \ce{EtOH} and centrifuged at \SI{10000}{rpm} for \SI{15}{\minute}. It was then allowed to stand for a \SI{10}{\minute}. The supernatant solution was removed and the black residue obtained was washed three times with \ce{EtOH} as above using centrifugation and one time with \ce{MeOH}. Each time, the colored or colorless supernatant solution was decanted. Finally, the product was transferred to a glass petri-dish using \ce{MeOH}, and kept overnight for drying at \SI{80}{\degreeCelsius} in a vacuum oven. The dried product thus obtained (CNT-PAMAM-G 2.0) was collected and weighed.
\subsection{Synthesis of higher generations of CNT-PAMAM}

The higher generations of the CNT-PAPAM were prepared by consecutive Michael addition and ester amidation reactions as described above.   Michael's addition with EDA and ester amidation with MA were repeated until the desired number of generations (up to generation 5.0) was achieved. While repeating the number of cycles, each product was thoroughly washed with ethanol and methanol several times. Finally,  the dendritic-modified CNT has been obtained. 
\subsection{Synthesis of Graphene Oxide-PAMAM (GO-PAMAM)}
Initially, GO was synthesized by the modified Hummers method\cite{shahriary2014graphene}. In brief, graphite (\SI{250}{\milli\gram}) and \ce{NaNO2} (\SI{250}{\milli\gram}) were mixed in \SI{12}{\milli\liter} \ce{H2SO4} (98\%) in an ice bath. After gradual addition of \ce{KMnO4} (\SI{1.5}{\gram}), \SI{23}{\milli\liter} deionized water was poured into the mixture steadily while keeping the temperature less than \SI{20}{\celsius} to prevent overheating. The ice bath was removed after 2 hours, and the mixture was stirred at \SI{35}{\celsius} for another 12 hours. Finally, it was treated with \SI{5}{\milli\liter} 30\% \ce{H2O2}. The resulting mixture was washed with \ce{HCl} and water by rinsing and centrifugation at 5000 rpm, and the product, GO, was dried under vacuum at room temperature.
\begin{figure}
    \centering
    \includegraphics[width=\linewidth]{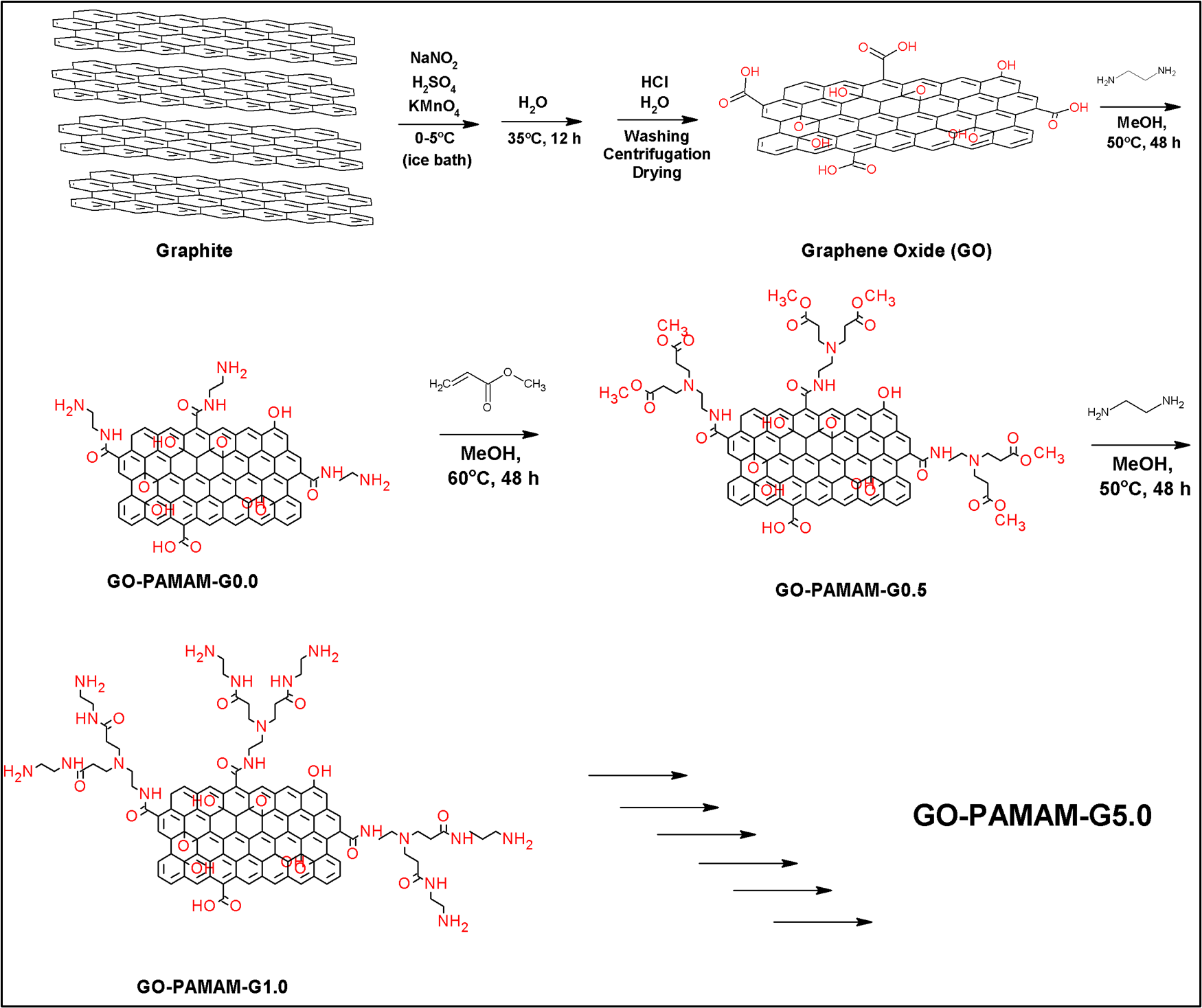}
    \caption{Reaction sequence of the de-novo synthesis of CNT-PAMAM-G5 from CNTReaction sequence of the de-novo synthesis of CNT-PAMAM-G5 from CNT}
    \label{fig:expt2}
\end{figure}
The \ce{-COOH} group present on the surface of GO was used as a linker for the divergent growth of PAMAM dendrons. A similar growth procedure was followed as used in the case of CNT-PAMAM. Thus, PAMAM dendrons starting from generation zero (G0) to generation five (G5) were grown on the surface of GO through the `grafting from' method by the divergent synthesis as described above for the synthesis of CNT-PAMAM in Section 1.2.  The overall reaction sequence for the synthesis of GO-PAMAM-G5 starting from graphite is depicted in Figure \label{fig:expt2}.

\subsection{Characterization of CNT-PAMAM and GO-PAMAM}

Each step of the above reactions was monitored by using Fourier Transform Infrared Spectroscopy (FTIR, make-JASCO, model-FT/IR-6300) with \SI{4}{cm^{-1}} resolution, range \SIrange{100}{4000}{cm^{-1}} and 16 numbers of scan in diffusion reflectance mode. The main features in the IR spectra of CNT-PAMAM and GO-PAMAM are discussed. The elemental analysis of the synthesized products was conducted in a CHNS analyzer (Make-Elemental, model- Vario Macro cube). Thermogravimetric analysis (TGA, Make: Netzsch, Model: STA 449 Jupiter F3) of the products was carried out to access the effect of temperature on them. The morphology and structure of the base nanomaterials compared with functionalized nanomaterials were performed in powder XRD (SmartLab powder X-ray diffractometer, Rigaku, Japan).

\subsection{Batch adsorption Studies}

A stock solution of \SI{1000}{\milli\gram\per\liter} of uranyl ions (\ce{UO2^2+}) was prepared by dissolving suitable amount of uranyl nitrate dihydrate salt in distilled water. The stock solution was then diluted to the required concentrations for performing adsorption studies. The pH of the solutions was modified by using \SI{0.1}{N} NaOH and \SI{0.1}{N} HCl. In each case, the adsorption capacity at equilibrium was calculated by using the following equation:

\begin{equation}
q_e = \frac{(C_o - C_e) \cdot V}{M}
\end{equation}

where $q_e$ (\si{\milli\gram\per\gram}) is the equilibrium adsorption capacity of the adsorbent, $C_o$ and $C_e$ are the \ce{UO2^2+} ion concentration (\si{\milli\gram\per\liter}) in the initial feed solution and at equilibrium, respectively, $V$ is the volume (\si{\liter}) of the solution taken initially and $M$ is the mass (\SI{}{\gram}) of the adsorbent added.

Trial studies were first conducted to calculate the time required to attain equilibrium, which was found to be 2 hours. For carrying out the adsorption experiments, \SIrange{5}{10}{\milli\gram} of CNT-PAMAM or GO-PAMAM was added to \SIrange{5}{10}{\milli\liter} of Hg (II) solutions of required concentrations and placed on an orbital shaker (make: IKA. model: KS 4000) for 3 hours. Separation of the \ce{UO2^2+} loaded CNT-PAMAM or GO-PAMAM was done by allowing the solution to settle down followed by centrifugation. Immediate analysis of the remaining solution was done using Inductively Coupled Plasma Optical Emission Spectroscopy ICP-OES (Horiba Scientific, Jobin Yvon Ultima 2). Experiments were performed by duplicate and average values were considered for the calculation. The effect of pH on the adsorption was first studied and it was found that maximum adsorption was found at pH 6 for both of the adsorbents of all full generation. Thus the equilibrium and other studies were conducted at pH 6. The initial uranyl ion concentration was varied from \SIrange{5}{1000}{\milli\gram\per\liter} with $V$ and $M$ fixed at \SI{5}{\milli\liter} and \SI{5}{\milli\gram}, respectively.

\bibliography{ref}